\documentclass{article}

\usepackage{microtype}
\usepackage{graphicx}
\usepackage{subfigure}
\usepackage{booktabs} 
\usepackage{kotex}
\usepackage{hyperref}

\usepackage[accepted]{icml2025}

\usepackage{amsmath}
\usepackage{amssymb}
\usepackage{mathtools}
\usepackage{amsthm}

\usepackage[capitalize,noabbrev]{cleveref}

\theoremstyle{plain}

\theoremstyle{definition}

\theoremstyle{remark}

\usepackage[textsize=tiny]{todonotes}
\usepackage{multirow}
\usepackage{makecell}
\usepackage{array}
\usepackage{caption}

\usepackage{bbm}
\usepackage{arydshln}
\usepackage{enumitem}

\icmltitlerunning{Accepted as Spotlight at ICML 2025 Workshop GenBio}

\begin{document}

\twocolumn[
\icmltitle{{DIVER-0 : A Fully Channel Equivariant EEG Foundation Model}}

\begin{icmlauthorlist}
\icmlauthor{Danny Dongyeop Han}{snu}
\icmlauthor{Ahhyun Lucy Lee}{snu}
\icmlauthor{Taeyang Lee}{snu}
\icmlauthor{Yonghyeon Gwon}{snu}
\icmlauthor{Sebin Lee}{snu}
\icmlauthor{Seongjin Lee}{snu}
\icmlauthor{David Keetae Park}{bnl}
\icmlauthor{Shinjae Yoo}{bnl}
\icmlauthor{Jiook Cha}{snu}
\icmlauthor{Chun Kee Chung}{snu}
\end{icmlauthorlist}

\icmlaffiliation{snu}{Seoul National University, Seoul, Republic of Korea}
\icmlaffiliation{bnl}{Brookhaven National Laboratory, New York, United States}

\icmlcorrespondingauthor{Jiook Cha}{cha.jiook@gmail.com}
\icmlcorrespondingauthor{Chun Kee Chung}{chungc@snu.ac.kr}

\icmlkeywords{Machine Learning, ICML}

\vskip 0.3in
]

\printAffiliationsAndNotice{}
\begin{abstract}

Electroencephalography (EEG) is a non-invasive technique widely used in brain-computer interfaces and clinical applications, yet existing EEG foundation models face limitations in modeling spatio-temporal brain dynamics and lack channel permutation equivariance, preventing robust generalization across diverse electrode configurations. To address these challenges, we propose DIVER-0, a novel EEG foundation model that demonstrates how full spatio-temporal attention—rather than segregated spatial or temporal processing—achieves superior performance when properly designed with Rotary Position Embedding (RoPE) for temporal relationships and binary attention biases for channel differentiation. We also introduce Sliding Temporal Conditional Positional Encoding (STCPE), which improves upon existing conditional positional encoding approaches by maintaining both temporal translation equivariance and channel permutation equivariance, enabling robust adaptation to arbitrary electrode configurations unseen during pretraining. Experimental results demonstrate that DIVER-0 achieves competitive performance with only 10\% of pretraining data while maintaining consistent results across all channel permutation conditions, validating its effectiveness for cross-dataset generalization and establishing key design principles for handling the inherent heterogeneity of neural recording setups.

\end{abstract}

\section{Introduction}
\label{introduction}

Electroencephalography (EEG) is a non-invasive neurophysiological technique that measures electrical activity generated by synchronized neuronal firing through electrodes placed on the scalp. With its high temporal resolution, portability, and affordability \cite{nicolas2012brain}, EEG has been widely and effectively utilized in brain-computer interfaces \cite{schalk2004bci2000,zhang2019cyborg}, clinical diagnostics \cite{meghdadi2021resting}, cognitive neuroscience research \cite{cohen2017does}, and in controlling assistive and rehabilitation devices for patients with neurological conditions \cite{meng2016noninvasive}.

EEG decoding has evolved significantly from early traditional machine learning approaches \cite{lotte2007review} that relied on hand-crafted features to sophisticated deep learning architectures \cite{craik2019eegdlreview}, including convolutional neural networks \cite{lawhern2018eegnet} and transformers \cite{song2022eegconformer} that automatically learn hierarchical representations from raw signals \cite{schirrmeister2017convnet, li2022ffcl, bagchi2022convtransformer, huang2022lrtcnn}. More recently, following the success of foundation models in other domains, researchers have begun embracing self-supervised learning (SSL) paradigms, pretraining large-scale models on massive amounts of unlabeled EEG data \cite{wang2024eegpt,jiang2024labram,foumani2024eeg2rep,shi2024fome,chien2022maeeg,kostas2021bendr,wang2024cbramod} to learn generic neural representations that can be fine-tuned for diverse downstream tasks. This paradigm shift promises to address fundamental challenges including data scarcity, cross-subject variability, and computational overhead of task-specific model development, while unlocking new capabilities for understanding and decoding complex neural signals.

However, existing EEG foundation models fail to address two critical limitations that constrain their effectiveness:

\textbf{Overly restrictive modeling of spatio-temporal brain dynamics}. 
Current approaches are constrained in its ability to capture the complex interactions between brain regions due to their treatment of multi-channel EEG data. Most methods fall into one of three restrictive categories. 
First, many approaches structurally segregate spatial and temporal processing by having separate spatial and temporal encoders \cite{chau2025popt,zhang2023brant,shi2024fome} or dedicated spatial and temporal attention blocks/heads \cite{dimofte2025cerebro,wang2024cbramod}. While computationally efficient, this rigid separation cannot adequately model the brain's distributed dynamics, where neural regions interact through temporally extended dependencies \cite{deco2011emerging} and temporal and phase lags \cite{stam2007phase,fries2005mechanism,varela2001brainweb}. Second, other methods collapse or flatten the channel dimension and eliminate spatial structure early in processing \cite{wang2024eegpt,foumani2024eeg2rep}. This creates an information bottleneck that prevents the model from querying specific spatial relationships in later layers.
Third, remaining approaches apply full attention with spatio-temporal input embeddings \cite{jiang2024labram}, which means that they cannot adapt to electrodes not seen during pretraining, which is common in EEG where different researchers use varying electrode montages \cite{jurcak200710, xu2020cross} or custom spatial configurations \cite{atcherson2007variability}, limiting applicability across diverse recording setups.

\textbf{Lack of channel permutation equivariance and poor generalization across electrode configurations}.
Current EEG foundation models fail to properly handle the fundamental requirement that model performance should be invariant to electrode ordering—a critical property for generalizing across diverse channel numbers and setups. Many methods use absolute channel embeddings that learn fixed representations for specific electrode positions, making them unable to generalize to unseen electrode configurations or montages. While recent methods like ACPE \cite{wang2024cbramod} attempted to address cross-dataset transfer by using asymmetric conditional positional encoding to dynamically learn spatial relationships, it still fails to maintain proper permutation equivariance because it applies convolutions across both spatial (channel) and temporal dimensions simultaneously. This architectural choice breaks the fundamental mathematical property that the model should perform identically regardless of channel ordering. 

To address these limitations, we propose a full-attention architecture with novel condition positional encoding schemes that maintain both temporal translation equivariance and channel permutation equivariance, enabling robust generalization across diverse electrode configurations and experimental setups.

\begin{itemize}[itemsep=0pt, topsep=0pt, parsep=0pt]
    \item We present \textbf{DIVER-0}, a new EEG foundation model that demonstrates competitive performance across representative downstream BCI tasks with different electrode configurations, while maintaining strict permutation equivariance—a critical property for cross-dataset generalization.

    \item We introduce the \textbf{DIVER transformer block} with unified spatio-temporal attention, combining Rotary Position Embedding (RoPE) \cite{su2024roformer} for temporal relationships and binary attention biases \cite{woo2024moirai} for channel differentiation while maintaining permutation equivariance across electrode orderings.
    \item We propose \textbf{Sliding Temporal Conditional Positional Encoding (STCPE)} that preserves both temporal translation equivariance and channel permutation equivariance, enabling robust generalization to arbitrary electrode configurations unseen during pretraining.
\end{itemize}

\begin{figure*}[h]
  \centering
  \includegraphics[width=0.9\textwidth]{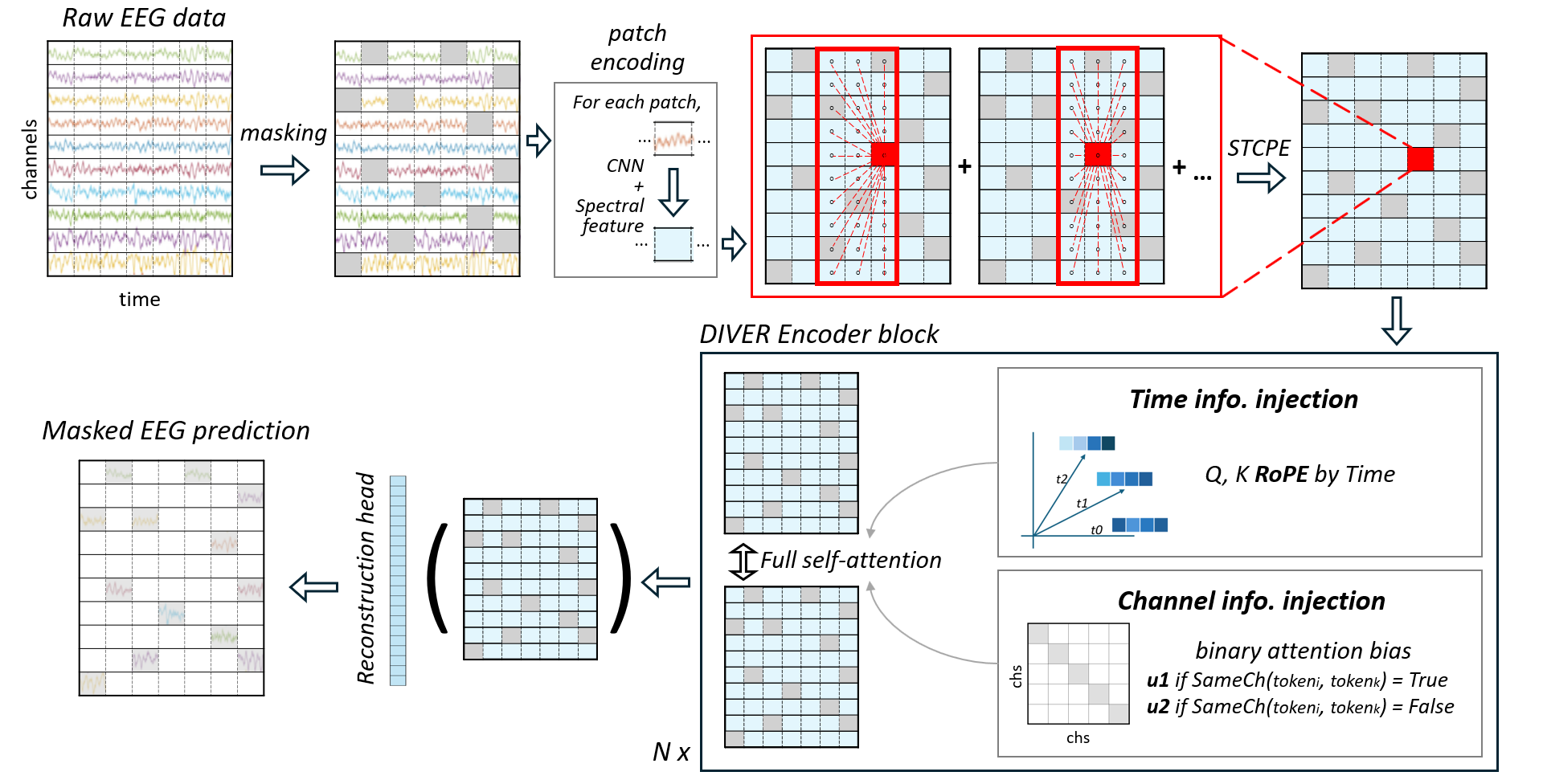}
 \caption{\textbf{Overview of DIVER-0 architecture and pretraining} EEG signals are segmented into 1-second patches (200 samples at 200 Hz). After masking, each patch undergoes patch encoding through patch-wise convolutional neural networks (CNNs) for temporal features, with spectral features obtained via Fast Fourier Transform (FFT) and element-wise addition. The Sliding Temporal Conditional Positional Embedding (STCPE) is computed by summing outputs from sliding transformer encoders across the temporal dimension. These embeddings are then processed by DIVER encoder blocks, which employ unified full self-attention across both temporal and spatial dimensions. Within each encoder block, temporal relationships are encoded using Rotary Position Embedding (RoPE) \cite{su2024roformer}, while channel identity is encoded through binary attention biases \cite{woo2024moirai,liu2024moiraimoe}. The model is pretrained using masked patch reconstruction.}
  \label{fig:wide_fig_top}
\end{figure*}

\section{Methods}
\label{methods}
\subsection{Architecture}

\subsubsection{Patch Encoding}\label{methods:patchencoding}
We divide the raw EEG signal from each channel into small patches of 1 second duration, with each patch containing 200 samples (at 200 Hz sampling rate). After masking, each patch undergoes a two-step patch encoding process. First, the raw patch is processed through patch-wise convolutional neural networks (CNNs) to extract temporal features. The spectral pathway then applies Fast Fourier Transform (FFT) to this CNN output. The final patch encoding is obtained by element-wise addition of the CNN features and their frequency domain transformation. This approach creates a unified patch representation that combines the original temporal features with their spectral characteristics derived from the same CNN-processed signal.

\subsubsection{Sliding Temporal Conditional Positional Embedding (STCPE)}\label{methods:STCPE}

Following patch encoding, we apply Sliding Temporal Conditional Positional Embedding (STCPE). Conditional Positional Encoding (CPE) originally uses convolutional networks to dynamically generate positional encodings from spatial-temporal neighborhoods, which was successfully adapted as Asymmetric Conditional Positional Encoding (ACPE) in CBraMod \cite{wang2024cbramod}, achieving state-of-the-art performance across multiple downstream BCI tasks. The STCPE mechanism, in contrast, operates by sliding a transformer block across the temporal dimension, where at each time step, the transformer processes patches from \textit{all channels simultaneously}, computing positional information by aggregating outputs across this temporal sliding window. The STCPE is constructed using DIVER encoder blocks (see Section~\ref{methods:DIVEReneocder}), which incorporate both Rotary Position Embedding (RoPE)\cite{su2024roformer} for temporal awareness and binary attention bias\cite{woo2024moirai} for channel differentiation, enabling the model to distinguish between different channels and timepoints during the positional encoding process.

Unlike ACPE \cite{wang2024cbramod} that applies asymmetric convolutions across both spatial (channel) and temporal dimensions simultaneously, our STCPE design maintains crucial mathematical properties: \textbf{temporal translation equivariance} and \textbf{channel-wise permutation equivariance}. This enables STCPE to provide generalizable positional encodings for arbitrary electrode orderings and configurations not seen during pretraining, while also being invariant to temporal shifts in the input signals.

To manage computational complexity, linear projection layers are applied before and after STCPE to reduce the embedding dimension during this positional encoding stage.
The final patch embeddings with positional information are obtained as $\mathbf{X}^o = \mathbf{X} + \mathbf{PE} = \{\mathbf{x}_{i,j} + \mathbf{pe}_{i,j} | i\in[1, 2, ...,C], j \in[1, 2, ...,N]\}$, where $\mathbf{X}^o \in \mathbb{R}^{C \times N \times D}$ represents the patch embeddings enhanced with relative positional information.

\subsubsection{DIVER encoder}\label{methods:DIVEReneocder}

After applying the STCPE, the sequence of embeddings is fed into the DIVER encoder, which follows a structure similar to the full attention Transformer enhanced with components from MOIRAI~\cite{woo2024moirai}. The encoder incorporates two key mechanisms: rotary positional embedding to encode temporal relationships between time points, and binary attention bias to inform the model when two electrodes represent the same channel. 

The attention mechanism can be formulated as follows, where for clarity we omit layer and attention head indices as well as scaling factors. The attention score between the $(i,m)$-th query (where $i$ denotes the time index and $m$ denotes the electrode index) and the $(j,n)$-th key is computed as:
\begin{equation}
    \begin{split}
        E_{ij,mn} &= (\mathbf{W}^Q\boldsymbol{x}_{i,m})^T \mathbf{R}_{i-j} (\mathbf{W}^K\boldsymbol{x}_{j,n}) \\
              &\quad + u^1 * \mathbbm{1}_{\{m=n\}} + u^2 * \mathbbm{1}_{\{m \neq n\}}, \\
        A_{ij,mn} &= \frac{\exp\{E_{ij,mn}\}}{\sum_{k,o} \exp\{E_{ik,mo}\}}
    \end{split}
    \label{eq:DIVER attention}
\end{equation}
where $\mathbf{W}^Q\boldsymbol{x}_{i,m},\mathbf{W}^K\boldsymbol{x}_{j,n} \in \mathbb{R}^{d_h}$ are the query and key vectors, $\mathbf{R}_{i-j} \in \mathbb{R}^{d_h \times d_h}$ is the rotary projection matrix, $u^1, u^2 \in \mathbb{R}$ are learnable scalars that can be different in each head, $\mathbbm{1}_{\{cond\}}=1$ if ${cond}$ is true, and $\mathbbm{1}_{\{cond\}}=0$ otherwise.

\subsection{Pretraining}

DIVER was pretrained on the extensive Temple University Hospital EEG Corpus (TUEG) dataset \cite{obeid2016tueg}, one of the largest publicly available EEG datasets comprising of 69,652 clinical recordings from 14,987 subjects. This dataset provides approximately 30,000 hours of EEG data, ideal for  pretraining models to learn robust neural representations across diverse clinical populations and pathological conditions.

The preprocessing pipeline closely followed the methodology established in CBraMod \cite{wang2024cbramod}. Briefly, the raw EEG signals underwent bandpass filtering (0.3-75 Hz) to remove low and high-frequency artifacts, followed by a 60 Hz notch filter to eliminate power line interference. All recordings were resampled to 200 Hz and segmented into 30-second non-overlapping training samples. Signals exceeding 100 μV amplitude were automatically rejected to ensure data quality, and 19 standard EEG channels following the 10-20 system were selected for analysis.

The DIVER backbone consists of 12 DIVER encoder blocks with a hidden dimension of 200, inner feed-forward dimensions of 800, and 10 attention heads. As illustrated in Figure~\ref{fig:wide_fig_top}, the model also employs a novel embedding strategy combining patch-wise CNN projection, spectral features, and Sliding Temporal Conditional Positional Encoding (STCPE) with a temporal window size of 7 seconds. 

Training was conducted using the AdamW optimizer with a learning rate of $5 \times 10^{-4}$ for pretraining on 10\% of the data and $1 \times 10^{-4}$ for full dataset pretraining, with corresponding weight decay values of $5 \times 10^{-4}$ and $1 \times 10^{-3}$ respectively. The model was implemented using Python 3.12.7 and PyTorch 2.5.1 with CUDA 12.1 support, and training was performed on compute nodes equipped with four NVIDIA A100 40GB GPUs.

\subsection{Finetuning}
Finetuning was done on two downstream datasets, FACED and PhysioNet-MI. 
FACED dataset \cite{chen2023faced} contains 32-channel EEG recordings from 123 subjects who watched 28 emotion-elicitation video clips covering nine emotion categories. PhysioNet-MI \cite{goldberger2000physionet} refers to the EEG Motor Movement/Imagery Dataset from PhysioNet \cite{goldberger2000physionet}, which consists of more than 1500 EEG recordings from 109 volunteers performing a series of real and imagined motor tasks, with data recorded from 64 electrodes. Implementation details including hyperparmeters are detailed in Appendix \ref{append_implementation_detail}.

\begin{table*}[t]
\footnotesize
\centering
\setlength{\tabcolsep}{4pt}
\caption{\textbf{Performance comparison on downstream EEG tasks}. Results are reported for emotion recognition (FACED, 9-class) and motor imagery classification (PhysioNet-MI, 4-class). Performance values for baseline methods are quoted from \cite{wang2024cbramod} using identical preprocessing and evaluation protocols. DIVER results are from our experiments. All metrics are reported as mean $\pm$ standard deviation across 5 random seeds. DIVER 10\% indicate that the model was pretrained using 10\% of the full TUEG dataset.}

\label{table1}
\begin{tabular}{lcccccc}
\toprule
 & \multicolumn{3}{c}{\textbf{FACED, 9-class}} & \multicolumn{3}{c}{\textbf{PhysioNet-MI, 4-class}} \\
\cmidrule(lr){2-4} \cmidrule(lr){5-7}
\textbf{Methods} & \textbf{Bal. Acc.(\%) } & \textbf{Kappa(\%)} & \textbf{F1(\%)} & \textbf{Bal. Acc. (\%) } & \textbf{Kappa(\%)} & \textbf{F1(\%)} \\
\midrule
EEGNet              
& 40.9 $\pm$ 1.2 & 33.4 $\pm$ 2.5 & 41.2 $\pm$ 1.4 
& 58.1 $\pm$ 1.3 & 44.7 $\pm$ 2.0 & 58.0 $\pm$ 1.2 \\
SPaRCNet            
& 46.7 $\pm$ 1.6 & 39.8 $\pm$ 2.9 & 47.3 $\pm$ 1.3 
& 59.3 $\pm$ 1.5 & 45.6 $\pm$ 2.3 & 59.4 $\pm$ 1.5 \\
ST-Transformer      
& 48.1 $\pm$ 0.8 & 41.4 $\pm$ 1.3 & 48.0 $\pm$ 1.0 
& 60.4 $\pm$ 0.8 & 47.1 $\pm$ 2.0 & 60.5 $\pm$ 0.8 \\
EEGConformer        
& 45.6 $\pm$ 1.3 & 38.6 $\pm$ 1.9 & 45.1 $\pm$ 1.1
& 60.5 $\pm$ 1.0 & 47.4 $\pm$ 1.7 & 60.6 $\pm$ 1.0 \\
\midrule
LaBraM-Base         
& 52.7 $\pm$ 1.1 & 47.0 $\pm$ 1.9 & 52.9 $\pm$ 1.0 
& 61.7 $\pm$ 1.2 & 49.1 $\pm$ 1.9 & 61.8 $\pm$ 1.4 \\
CBraMod             
& \underline{55.1 $\pm$ 0.9} & \underline{50.4 $\pm$ 1.2} & \underline{56.2 $\pm$ 0.9} 
& \textbf{64.2 $\pm$ 0.9} & \textbf{52.2 $\pm$ 1.7} & \textbf{64.3 $\pm$ 1.0} \\
\midrule
\textbf{DIVER 10\% (Ours)} 
& \textbf{59.2 $\pm$ 0.8} & \textbf{54.0 $\pm$ 0.9} & \textbf{59.6 $\pm$ 0.7} 
& \underline{62.8 $\pm$ 0.5} & \underline{50.4 $\pm$ 0.7} & \underline{62.9 $\pm$ 0.5} \\
\bottomrule
\end{tabular}
\end{table*}

\begin{table*}[t]
\centering
\footnotesize
\setlength{\tabcolsep}{4pt}
\caption{\textbf{Ablation analysis of pretrained model components.} Results are reported for emotion recognition (FACED, 9-class) and motor imagery classification (PhysioNet-MI, 4-class). DIVER 10\% and CBraMod 10\% indicate that the models were pretrained using 10\% of the full TUEG dataset. All metrics are reported as mean $\pm$ standard deviation across 5 random seeds.}
\label{table2}
\begin{tabular}{lcccccc}
\toprule
 & \multicolumn{3}{c}{\textbf{FACED, 9-class}} & \multicolumn{3}{c}{\textbf{PhysioNet-MI, 4-class}}\\ 
\cmidrule(lr){2-4} \cmidrule(lr){5-7}
\textbf{Methods} & \textbf{Bal. Acc. (\%)} & \textbf{Kappa(\%)} & \textbf{F1(\%)} & \textbf{Bal. Acc. (\%)} & \textbf{Kappa(\%)} & \textbf{F1(\%)} \\
\midrule
\textbf{DIVER 10\% (Ours)} 
& \textbf{ 59.2 $\pm$ 0.8 } & \textbf{ 54.0 $\pm$ 0.9} & \textbf{59.6 $\pm$ 0.7}
& 62.8 $\pm$ 0.5 & 50.4 $\pm$ 0.7 & 62.9 $\pm$ 0.5 \\
 \quad w/o patch-wise CNN encoding   
& 57.3 $\pm$ 0.9 & 51.7 $\pm$ 1.0 & 57.4 $\pm$ 0.9 
& 61.9 $\pm$ 0.5 & 49.3 $\pm$ 0.6 & 62.1 $\pm$ 0.5 \\
 \quad  w/o spectral embedding       
& 58.0 $\pm$ 1.1 & 52.6 $\pm$ 1.3 & 58.3 $\pm$ 1.2 
& \textbf{63.4 $\pm$ 0.5} & \textbf{51.1 $\pm$ 0.6} & \textbf{63.5 $\pm$ 0.5} \\
 \quad w/o STCPE            
& 58.4 $\pm$ 1.0 & 53.0 $\pm$ 1.1 & 58.6 $\pm$ 1.0 
& 62.8 $\pm$ 0.5 & 50.5 $\pm$ 0.6 & 63.0 $\pm$ 0.5 \\
  \quad STCPE $\rightarrow$ ACPE                 
& \underline{58.7 $\pm$ 0.9} & \underline{53.4 $\pm$ 1.1} & \underline{59.0 $\pm$ 1.0} 
& 62.4 $\pm$ 0.3 & 49.9 $\pm$ 0.4 & 62.5 $\pm$ 0.3 \\
\hdashline
  \quad w/o RoPE             
& 57.7 $\pm$ 0.6 & 52.2 $\pm$ 0.7 & 58.1 $\pm$ 0.6 
& 62.6 $\pm$ 0.4 & 50.2 $\pm$ 0.5 & 62.8 $\pm$ 0.4 \\
  \quad w/o Binary attention bias      
& 58.1 $\pm$ 1.0 & 52.6 $\pm$ 1.1 & 58.4 $\pm$ 1.0 
& 62.8 $\pm$ 0.5 & 50.4 $\pm$ 0.7 & 63.0 $\pm$ 0.6 \\
  \quad DIVER $\rightarrow$ Vanilla block
& 55.3 $\pm$ 1.9 & 49.5 $\pm$ 2.2 & 55.5 $\pm$ 1.9 
& 61.6 $\pm$ 1.5 & 48.8 $\pm$ 1.9 & 61.7 $\pm$ 1.5 \\
  \quad DIVER $\rightarrow$ CBraMod block
& 57.0 $\pm$ 0.6 & 51.5 $\pm$ 0.6 & 57.4 $\pm$ 0.5 
& \underline{63.1 $\pm$ 0.8} & \underline{50.8 $\pm$ 1.0} & \underline{63.2 $\pm$ 0.8} \\
\midrule
CBraMod 10\%   
& 56.5 $\pm$ 0.8 & 51.0 $\pm$ 1.0 & 56.9 $\pm$ 0.8 
& 62.4 $\pm$ 0.6 & 49.9 $\pm$ 0.8 & 62.6 $\pm$ 0.7\\
\bottomrule
\end{tabular}
\end{table*}

\begin{table*}[t]
\centering
\footnotesize
\caption{Performance comparison across Pretraining $\times$ Finetuning configurations on the FACED dataset. Balanced Accuracy, Kappa, and F1 are reported. ('w/o' denotes our model without the corresponding component, and '$\rightarrow$' denotes replacement of our model's component with the alternative component.) }
\label{table3}
\renewcommand{\arraystretch}{1.2}
\setlength{\tabcolsep}{4pt}

\begin{tabular}{cc|l|ccc|ccc}
\toprule
\multicolumn{3}{c}{} & \multicolumn{6}{c}{\textbf{Pretrain}} \\
\multicolumn{3}{c}{} & \multicolumn{3}{c}{\textbf{Intact}} & \multicolumn{3}{c}{\textbf{Permute}} \\
\midrule
\multicolumn{2}{c|}{} & \textbf{Methods} & \textbf{Bal. Acc.(\%)} & \textbf{Kappa(\%)} & \textbf{F1(\%)} 
& \textbf{Bal. Acc.(\%)} & \textbf{Kappa(\%)} & \textbf{F1(\%)} \\
\cmidrule{3-9}
\multirow[c]{14}{*}{\rotatebox[origin=c]{90}{\textbf{Finetuning}}}
& \multirow{7}{*}{\textbf{Intact}}
& \textbf{DIVER 10\% (Ours)} 
& \textbf{59.2 $\pm$ 0.9} & \textbf{54.0 $\pm$ 0.9} & \textbf{59.6 $\pm$ 0.7} 
& \underline{59.6 $\pm$ 0.8} & \underline{54.4 $\pm$ 0.9} & \underline{59.9 $\pm$ 0.8} \\
& & \quad w/o STCPE
& \underline{58.4 $\pm$ 1.0} & \underline{53.0 $\pm$ 1.1} & \underline{58.6 $\pm$ 1.0}
& 58.4 $\pm$ 0.8 & 52.9 $\pm$ 0.9 & 58.5 $\pm$ 0.8 \\
& & \quad STCPE $\rightarrow$ ACPE 
& 56.7 $\pm$ 1.1 & 51.2 $\pm$ 1.3 & 57.2 $\pm$ 1.1 
& \textbf{60.1 $\pm$ 0.9} & \textbf{54.8 $\pm$ 1.1} & \textbf{60.3 $\pm$ 0.9} \\
& & \quad w/o Binary attention bias
& 58.1 $\pm$ 1.0 & 52.6 $\pm$ 1.1 & 58.4 $\pm$ 1.0  
& 58.1 $\pm$ 1.3 & 52.7 $\pm$ 1.4 & 58.5 $\pm$ 1.2\\
& & \quad DIVER $\rightarrow$ Vanilla block
& 55.3 $\pm$ 1.9 & 49.5 $\pm$ 2.2 & 55.5 $\pm$ 1.9 
& 56.3 $\pm$ 0.7 & 50.7 $\pm$ 0.9 & 56.6 $\pm$ 0.8 \\
& & \quad DIVER $\rightarrow$ CBraMod block
& 57.0 $\pm$ 0.6 & 51.5 $\pm$ 0.6 & 57.4 $\pm$ 0.5 
& 56.8 $\pm$ 0.6 & 51.2 $\pm$ 0.7 & 57.1 $\pm$ 0.6 \\
\cline{3-9}
& & CBraMod 10\%
& 56.5 $\pm$ 0.8 & 51.0 $\pm$ 1.0 & 56.9 $\pm$ 0.8 
& 58.7 $\pm$ 0.4 & 53.4 $\pm$ 0.4 & 59.0 $\pm$ 0.5 \\
\cmidrule{2-9}

& \multirow{7}{*}{\textbf{Permute}}
& \textbf{DIVER 10\% (Ours)} 
& \underline{59.1 $\pm$ 0.4} & \underline{53.8 $\pm$ 0.5} & \underline{59.4 $\pm$ 0.3} 
& \underline{59.0 $\pm$ 0.7} & \underline{53.6 $\pm$ 0.8} & \textbf{59.3 $\pm$ 0.7} \\
& & \quad w/o STCPE 
& 58.7 $\pm$ 0.4 & 53.3 $\pm$ 0.4 & 58.9 $\pm$ 0.3   
& \textbf{59.1 $\pm$ 0.4} & \textbf{53.6 $\pm$ 0.4} & 58.9 $\pm$ 0.2 \\
& & \quad STCPE $\rightarrow$ ACPE 
& \textbf{60.1 $\pm$ 0.9} & \textbf{54.8 $\pm$ 1.0} & \textbf{60.3 $\pm$ 0.8}
& 58.8 $\pm$ 0.4 & 53.4 $\pm$ 0.5 & \underline{59.1 $\pm$ 0.4} \\
& & \quad w/o Binary attention bias  
& 58.6 $\pm$ 1.2 & 53.2 $\pm$ 1.4 & 58.9 $\pm$ 1.3 
& 58.7 $\pm$ 0.2 & 53.3 $\pm$ 0.2 & 59.0 $\pm$ 0.2 \\
& & \quad DIVER $\rightarrow$ Vanilla block
& 57.6 $\pm$ 1.7 & 51.8 $\pm$ 1.8 & 57.9 $\pm$ 0.2 
& 55.6 $\pm$ 0.8 & 49.9 $\pm$ 0.9 & 55.9 $\pm$ 0.7 \\
& & \quad DIVER $\rightarrow$ CBraMod block
& 57.0 $\pm$ 0.8 & 51.4 $\pm$ 0.9 & 57.3 $\pm$ 0.9  
& 57.1 $\pm$ 0.6 & 51.6 $\pm$ 0.7 & 57.4 $\pm$ 0.6 \\
\cline{3-9}
& & CBraMod 10\%
& 56.5 $\pm$ 0.5 & 50.9 $\pm$ 0.6 & 56.8 $\pm$ 0.5
& 58.0 $\pm$ 0.9 & 52.5 $\pm$ 1.1 & 58.2 $\pm$ 1.0 \\

\bottomrule
\end{tabular}
\end{table*}

\section{Experiments}
\label{results}

\subsection{Downstream Task Performance}

We evaluate DIVER-0 on two representative BCI tasks, FACED~\cite{chen2023faced} and PhysioNet-MI~\cite{goldberger2000physionet}, with different electrode configurations. Table~\ref{table1} presents the performance comparison against state-of-the-art baselines including both non-foundation models (EEGNet, SPaRCNet, ST-Transformer, EEGConformer) and recent EEG foundation models (LaBraM-Base, CBraMod).

DIVER-0 pretrained on 10\% of the TUEG corpus achieves competitive performance across both tasks. On FACED, DIVER-0 achieves 59.2\% balanced accuracy and 54.0\% Cohen's Kappa, outperforming all baselines including CBraMod (55.1\% balanced accuracy). On PhysioNet-MI, DIVER-0 achieves 62.8\% balanced accuracy and 50.4\% Cohen's Kappa, demonstrating robust cross-dataset transfer. While CBraMod achieves slightly higher performance (64.2\%), DIVER-0's competitive results using only 10\% of pretraining data highlight the efficiency of our approach.

These results validate the effectiveness of our unified spatio-temporal attention approach and STCPE for handling diverse electrode configurations while maintaining strict permutation equivariance.

\subsection{Component Ablation Analysis}
\label{subsec: component ablation analysis}

To validate the contribution of each architectural component, we conduct comprehensive ablation studies presented in Table~\ref{table2}. For computational efficiency, all models in this analysis were pretrained using 10\% of the TUEG corpus. We systematically remove or replace key components: patch-wise CNN encoding, spectral embedding, STCPE, RoPE, binary attention biases, and compare different transformer block designs.

The results reveal nuanced insights about component contributions across different task types. On the FACED emotion recognition task, DIVER-0 achieves the highest performance (59.2\% balanced accuracy), with each component contributing meaningfully: removing patch-wise CNN encoding leads to a 1.9\% drop, removing spectral embedding causes a 1.2\% decrease, and replacing DIVER blocks with vanilla transformers results in a substantial 3.9\% performance degradation. This validates the effectiveness of our unified spatio-temporal attention for complex, distributed brain processes involved in emotion recognition.

Interestingly, on the PhysioNet-MI motor imagery task, while DIVER-0 outperforms CBraMod 10\% (62.8\% vs 62.4\%), some ablated configurations achieve slightly higher performance. Specifically, removing spectral embedding yields 63.4\% balanced accuracy, and replacing DIVER blocks with CBraMod blocks achieves 63.1\%.

This pattern suggests that motor imagery classification, which involves localized cortical sources around the sensorimotor cortex \cite{hameed2025enhancing, smith2014non}, may benefit from more constrained attention patterns that focus on specific spatial-temporal relationships rather than full spatio-temporal interactions. This contrasts with emotion processing, which involves distributed cortico-limbic circuits encompassing deep subcortical structures such as the amygdala, hippocampus, and anterior insula \cite{pessoa2017network}, whose widespread cortical manifestations in EEG activity benefit from our model's ability to capture long-range spatio-temporal dependencies \cite{liu2023eeg, valderrama2025identifying}. In fact, motor imagery BCI is primarily confined to accessible sensorimotor cortical areas \cite{kim2025neurophysiological}, where more focused attention mechanisms may be sufficient. The relatively small dataset size of PhysioNet-MI may also favor the reduced parameter complexity of ablated configurations over the full model's capacity to capture complex distributed interactions.

These results highlight an important insight: while unified spatio-temporal attention excels for distributed brain processes (emotion recognition), more focused attention mechanisms may be advantageous for spatially localized tasks (motor imagery), particularly with limited training data. Nevertheless, DIVER-0 demonstrates competitive performance across both task types while maintaining the crucial advantage of permutation equivariance for cross-dataset generalization.

\subsection{Channel Permutation Analysis}
\label{subsec: channel permute}

A key advantage of DIVER-0 is its permutation equivariance with respect to channel ordering. Table~\ref{table3} presents a systematic analysis where we evaluate models under four conditions: intact pretraining with intact finetuning, intact pretraining with permuted finetuning, permuted pretraining with intact finetuning, and permuted pretraining with permuted finetuning. To isolate the encoder's robustness, our permutation protocol involves permuting the input channels, processing them through the encoder, then permuting them back to the original order before the classifier, ensuring we specifically test the underlying model's channel permutation robustness rather than classifier adaptation.

DIVER-0 demonstrates remarkable robustness across all permutation conditions, with balanced accuracy remaining virtually unchanged (ranging from 59.0\% to 59.6\%) regardless of channel ordering during pretraining or finetuning. This stability validates our claim about permutation equivariance and demonstrates practical advantages for real-world deployment where electrode montages may vary across research groups and clinical settings.

Interestingly, when STCPE is replaced with ACPE, we observe a nuanced pattern: ACPE performs well when either pretraining or finetuning involves permutation, but shows degraded performance when both are intact or both are permuted. We hypothesize this occurs because ACPE's convolutional design can adapt to become agnostic to channel orderings during permuted pretraining, learning spatially-invariant representations. Subsequently, during intact finetuning, ACPE can leverage its inherent channel discrimination capabilities to exploit specific spatial relationships.

This analysis reveals an important trade-off: while our current STCPE achieves robust permutation equivariance through binary channel differentiation (distinguishing same-channel from cross-channel interactions), it sacrifices detailed channel-specific discrimination (the ability to identify and leverage individual electrode characteristics). Future work could explore injecting explicit channel position information alongside binary attention biases to maintain permutation equivariance while enabling finer-grained spatial awareness. Nevertheless, DIVER-0's consistent performance across all permutation conditions demonstrates the practical value of our approach for robust cross-dataset generalization.

\section{Conclusion}\label{Conclusion}

We present DIVER-0, a novel EEG foundation model that addresses critical limitations in existing approaches through two key innovations. First, we introduce the DIVER transformer block with unified spatio-temporal attention, combining Rotary Position Embedding (RoPE) for temporal relationships and binary attention biases for channel differentiation. Second, we propose Sliding Temporal Conditional Positional Encoding (STCPE), which maintains both temporal translation equivariance and channel permutation equivariance. 

Our experiments demonstrate that DIVER-0 achieves competitive performance across representative BCI tasks while maintaining strict permutation equivariance, enabling robust generalization to unseen electrode configurations. Notably, even when pretrained on only 10\% of the TUEG corpus, DIVER-0 outperforms existing foundation models on emotion recognition tasks, highlighting the effectiveness of our unified spatio-temporal attention approach for distributed brain processes.

Furthermore, the consistent performance of DIVER-0 across all channel permutation conditions demonstrates the practical value of maintaining strict permutation equivariance for cross-dataset generalization. Our unified spatio-temporal attention approach successfully captures complex neural dynamics while preserving the mathematical properties essential for robust deployment across diverse electrode configurations.

\textbf{Future Work.} Several directions warrant investigation: (1) Rigorous evaluation across more diverse datasets and tasks to systematically validate our hypothesis that unified spatio-temporal attention benefits tasks involving distributed brain networks (e.g., emotion, cognition) while spatially constrained attention may be more suitable for localized cortical processes (e.g., sensorimotor functions); (2) Developing methods to incorporate explicit channel position information while maintaining permutation equivariance, potentially through relative biases or unified topology mapping approaches as explored in MMM~\citep{yi2023learningMMM}; (3) Scaling to larger datasets (full TUEG and additional EEG corpora) and model sizes to explore the limits of EEG foundation model capabilities; (4) Systematic analysis of why ACPE combined with channel permutation shows promising results in certain conditions, and whether this transfers to other downstream applications.

Our work demonstrates the importance of inherently modeling both spatial and temporal EEG dynamics through unified attention mechanisms with embedded channel differentiation via binary attention biases. The success of full spatio-temporal attention for distributed brain processes, combined with the critical requirement for permutation equivariance, establishes key design principles for future EEG foundation models. DIVER-0 thus provides a robust foundation for handling the inherent heterogeneity of neural recording setups while maintaining mathematical guarantees essential for cross-dataset generalization.

\nocite{langley00}


\begin{thebibliography}{45}
\providecommand{\natexlab}[1]{#1}
\providecommand{\url}[1]{\texttt{#1}}
\expandafter\ifx\csname urlstyle\endcsname\relax
  \providecommand{\doi}[1]{doi: #1}\else
  \providecommand{\doi}{doi: \begingroup \urlstyle{rm}\Url}\fi

\bibitem[Atcherson et~al.(2007)Atcherson, Gould, Pousson, and Prout]{atcherson2007variability}
Atcherson, S.~R., Gould, H.~J., Pousson, M.~A., and Prout, T.~M.
\newblock Variability of electrode positions using electrode caps.
\newblock \emph{Brain topography}, 20:\penalty0 105--111, 2007.

\bibitem[Bagchi \& Bathula(2022)Bagchi and Bathula]{bagchi2022convtransformer}
Bagchi, S. and Bathula, D.~R.
\newblock Eeg-convtransformer for single-trial eeg-based visual stimulus classification.
\newblock \emph{Pattern Recognition}, 129:\penalty0 108757, 2022.

\bibitem[Chau et~al.(2025)Chau, Wang, Talukder, Subramaniam, Soedarmadji, Yue, Katz, and Barbu]{chau2025popt}
Chau, G., Wang, C., Talukder, S., Subramaniam, V., Soedarmadji, S., Yue, Y., Katz, B., and Barbu, A.
\newblock Population transformer: Learning population-level representations of neural activity.
\newblock \emph{ArXiv}, pp.\  arXiv--2406, 2025.

\bibitem[Chen et~al.(2023)Chen, Wang, Huang, Hu, Shen, and Zhang]{chen2023faced}
Chen, J., Wang, X., Huang, C., Hu, X., Shen, X., and Zhang, D.
\newblock A large finer-grained affective computing eeg dataset.
\newblock \emph{Scientific Data}, 10\penalty0 (1):\penalty0 740, 2023.

\bibitem[Chien et~al.(2022)Chien, Goh, Sandino, and Cheng]{chien2022maeeg}
Chien, H.-Y.~S., Goh, H., Sandino, C.~M., and Cheng, J.~Y.
\newblock Maeeg: Masked auto-encoder for eeg representation learning.
\newblock \emph{arXiv preprint arXiv:2211.02625}, 2022.

\bibitem[Cohen(2017)]{cohen2017does}
Cohen, M.~X.
\newblock Where does eeg come from and what does it mean?
\newblock \emph{Trends in neurosciences}, 40\penalty0 (4):\penalty0 208--218, 2017.

\bibitem[Craik et~al.(2019)Craik, He, and Contreras-Vidal]{craik2019eegdlreview}
Craik, A., He, Y., and Contreras-Vidal, J.~L.
\newblock Deep learning for electroencephalogram (eeg) classification tasks: a review.
\newblock \emph{Journal of neural engineering}, 16\penalty0 (3):\penalty0 031001, 2019.

\bibitem[Deco et~al.(2011)Deco, Jirsa, and McIntosh]{deco2011emerging}
Deco, G., Jirsa, V.~K., and McIntosh, A.~R.
\newblock Emerging concepts for the dynamical organization of resting-state activity in the brain.
\newblock \emph{Nature reviews neuroscience}, 12\penalty0 (1):\penalty0 43--56, 2011.

\bibitem[Dimofte et~al.(2025)Dimofte, Bucagu, Ingolfsson, Wang, Cossettini, Benini, and Li]{dimofte2025cerebro}
Dimofte, A., Bucagu, G.~A., Ingolfsson, T.~M., Wang, X., Cossettini, A., Benini, L., and Li, Y.
\newblock Cerebro: Compact encoder for representations of brain oscillations using efficient alternating attention.
\newblock \emph{arXiv preprint arXiv:2501.10885}, 2025.

\bibitem[Fries(2005)]{fries2005mechanism}
Fries, P.
\newblock A mechanism for cognitive dynamics: neuronal communication through neuronal coherence.
\newblock \emph{Trends in cognitive sciences}, 9\penalty0 (10):\penalty0 474--480, 2005.

\bibitem[Goldberger et~al.(2000)Goldberger, Amaral, Glass, Hausdorff, Ivanov, Mark, Mietus, Moody, Peng, and Stanley]{goldberger2000physionet}
Goldberger, A.~L., Amaral, L.~A., Glass, L., Hausdorff, J.~M., Ivanov, P.~C., Mark, R.~G., Mietus, J.~E., Moody, G.~B., Peng, C.-K., and Stanley, H.~E.
\newblock Physiobank, physiotoolkit, and physionet: components of a new research resource for complex physiologic signals.
\newblock \emph{circulation}, 101\penalty0 (23):\penalty0 e215--e220, 2000.

\bibitem[Hameed et~al.(2025)Hameed, Khan, Ahmed, Aftab, and Fazal]{hameed2025enhancing}
Hameed, I., Khan, D.~M., Ahmed, S.~M., Aftab, S.~S., and Fazal, H.
\newblock Enhancing motor imagery eeg signal decoding through machine learning: A systematic review of recent progress.
\newblock \emph{Computers in Biology and Medicine}, 185:\penalty0 109534, 2025.

\bibitem[Huang et~al.(2022)Huang, Chang, Yan, Yang, Luo, and Pei]{huang2022lrtcnn}
Huang, W., Chang, W., Yan, G., Yang, Z., Luo, H., and Pei, H.
\newblock Eeg-based motor imagery classification using convolutional neural networks with local reparameterization trick.
\newblock \emph{Expert Systems with Applications}, 187:\penalty0 115968, 2022.

\bibitem[Jiang et~al.(2024)Jiang, Zhao, and Lu]{jiang2024labram}
Jiang, W.-B., Zhao, L.-M., and Lu, B.-L.
\newblock Large brain model for learning generic representations with tremendous eeg data in bci.
\newblock \emph{arXiv preprint arXiv:2405.18765}, 2024.

\bibitem[Jurcak et~al.(2007)Jurcak, Tsuzuki, and Dan]{jurcak200710}
Jurcak, V., Tsuzuki, D., and Dan, I.
\newblock 10/20, 10/10, and 10/5 systems revisited: their validity as relative head-surface-based positioning systems.
\newblock \emph{Neuroimage}, 34\penalty0 (4):\penalty0 1600--1611, 2007.

\bibitem[Kim et~al.(2025)Kim, Kim, and Lee]{kim2025neurophysiological}
Kim, S.-H., Kim, S.-J., and Lee, D.-H.
\newblock Neurophysiological analysis in motor and sensory cortices for improving motor imagination.
\newblock In \emph{2025 13th International Conference on Brain-Computer Interface (BCI)}, pp.\  1--4. IEEE, 2025.

\bibitem[Kostas et~al.(2021)Kostas, Aroca-Ouellette, and Rudzicz]{kostas2021bendr}
Kostas, D., Aroca-Ouellette, S., and Rudzicz, F.
\newblock Bendr: Using transformers and a contrastive self-supervised learning task to learn from massive amounts of eeg data.
\newblock \emph{Frontiers in Human Neuroscience}, 15:\penalty0 653659, 2021.

\bibitem[Langley(2000)]{langley00}
Langley, P.
\newblock Crafting papers on machine learning.
\newblock In Langley, P. (ed.), \emph{Proceedings of the 17th International Conference on Machine Learning (ICML 2000)}, pp.\  1207--1216, Stanford, CA, 2000. Morgan Kaufmann.

\bibitem[Lawhern et~al.(2018)Lawhern, Solon, Waytowich, Gordon, Hung, and Lance]{lawhern2018eegnet}
Lawhern, V.~J., Solon, A.~J., Waytowich, N.~R., Gordon, S.~M., Hung, C.~P., and Lance, B.~J.
\newblock Eegnet: a compact convolutional neural network for eeg-based brain--computer interfaces.
\newblock \emph{Journal of neural engineering}, 15\penalty0 (5):\penalty0 056013, 2018.

\bibitem[Li et~al.(2022)Li, Ding, Zhang, and Xiu]{li2022ffcl}
Li, H., Ding, M., Zhang, R., and Xiu, C.
\newblock Motor imagery eeg classification algorithm based on cnn-lstm feature fusion network.
\newblock \emph{Biomedical signal processing and control}, 72:\penalty0 103342, 2022.

\bibitem[Liu et~al.(2023)Liu, Hu, Shen, Lv, Song, and Zhang]{liu2023eeg}
Liu, J., Hu, X., Shen, X., Lv, Z., Song, S., and Zhang, D.
\newblock The eeg microstate representation of discrete emotions.
\newblock \emph{International Journal of Psychophysiology}, 186:\penalty0 33--41, 2023.

\bibitem[Liu et~al.(2024)Liu, Liu, Woo, Aksu, Liang, Zimmermann, Liu, Savarese, Xiong, and Sahoo]{liu2024moiraimoe}
Liu, X., Liu, J., Woo, G., Aksu, T., Liang, Y., Zimmermann, R., Liu, C., Savarese, S., Xiong, C., and Sahoo, D.
\newblock Moirai-moe: Empowering time series foundation models with sparse mixture of experts.
\newblock \emph{arXiv preprint arXiv:2410.10469}, 2024.

\bibitem[Lotte et~al.(2007)Lotte, Congedo, L{\'e}cuyer, Lamarche, and Arnaldi]{lotte2007review}
Lotte, F., Congedo, M., L{\'e}cuyer, A., Lamarche, F., and Arnaldi, B.
\newblock A review of classification algorithms for eeg-based brain--computer interfaces.
\newblock \emph{Journal of neural engineering}, 4\penalty0 (2):\penalty0 R1, 2007.

\bibitem[Meghdadi et~al.(2021)Meghdadi, Stevanovi{\'c}~Kari{\'c}, McConnell, Rupp, Richard, Hamilton, Salat, and Berka]{meghdadi2021resting}
Meghdadi, A.~H., Stevanovi{\'c}~Kari{\'c}, M., McConnell, M., Rupp, G., Richard, C., Hamilton, J., Salat, D., and Berka, C.
\newblock Resting state eeg biomarkers of cognitive decline associated with alzheimer’s disease and mild cognitive impairment.
\newblock \emph{PloS one}, 16\penalty0 (2):\penalty0 e0244180, 2021.

\bibitem[Meng et~al.(2016)Meng, Zhang, Bekyo, Olsoe, Baxter, and He]{meng2016noninvasive}
Meng, J., Zhang, S., Bekyo, A., Olsoe, J., Baxter, B., and He, B.
\newblock Noninvasive electroencephalogram based control of a robotic arm for reach and grasp tasks.
\newblock \emph{Scientific Reports}, 6\penalty0 (1):\penalty0 38565, 2016.

\bibitem[Mohammadi~Foumani et~al.(2024)Mohammadi~Foumani, Mackellar, Ghane, Irtza, Nguyen, and Salehi]{foumani2024eeg2rep}
Mohammadi~Foumani, N., Mackellar, G., Ghane, S., Irtza, S., Nguyen, N., and Salehi, M.
\newblock Eeg2rep: enhancing self-supervised eeg representation through informative masked inputs.
\newblock In \emph{Proceedings of the 30th ACM SIGKDD Conference on Knowledge Discovery and Data Mining}, pp.\  5544--5555, 2024.

\bibitem[Nicolas-Alonso \& Gomez-Gil(2012)Nicolas-Alonso and Gomez-Gil]{nicolas2012brain}
Nicolas-Alonso, L.~F. and Gomez-Gil, J.
\newblock Brain computer interfaces, a review.
\newblock \emph{sensors}, 12\penalty0 (2):\penalty0 1211--1279, 2012.

\bibitem[Obeid \& Picone(2016)Obeid and Picone]{obeid2016tueg}
Obeid, I. and Picone, J.
\newblock The temple university hospital eeg data corpus.
\newblock \emph{Frontiers in neuroscience}, 10:\penalty0 196, 2016.

\bibitem[Pessoa(2017)]{pessoa2017network}
Pessoa, L.
\newblock A network model of the emotional brain.
\newblock \emph{Trends in cognitive sciences}, 21\penalty0 (5):\penalty0 357--371, 2017.

\bibitem[Schalk et~al.(2004)Schalk, McFarland, Hinterberger, Birbaumer, and Wolpaw]{schalk2004bci2000}
Schalk, G., McFarland, D.~J., Hinterberger, T., Birbaumer, N., and Wolpaw, J.~R.
\newblock Bci2000: a general-purpose brain-computer interface (bci) system.
\newblock \emph{IEEE Transactions on biomedical engineering}, 51\penalty0 (6):\penalty0 1034--1043, 2004.

\bibitem[Schirrmeister et~al.(2017)Schirrmeister, Springenberg, Fiederer, Glasstetter, Eggensperger, Tangermann, Hutter, Burgard, and Ball]{schirrmeister2017convnet}
Schirrmeister, R.~T., Springenberg, J.~T., Fiederer, L. D.~J., Glasstetter, M., Eggensperger, K., Tangermann, M., Hutter, F., Burgard, W., and Ball, T.
\newblock Deep learning with convolutional neural networks for eeg decoding and visualization.
\newblock \emph{Human brain mapping}, 38\penalty0 (11):\penalty0 5391--5420, 2017.

\bibitem[Shi et~al.(2024)Shi, Zhao, Yuan, Wang, Hu, Yu, and Zhang]{shi2024fome}
Shi, E., Zhao, K., Yuan, Q., Wang, J., Hu, H., Yu, S., and Zhang, S.
\newblock Fome: A foundation model for eeg using adaptive temporal-lateral attention scaling.
\newblock \emph{arXiv preprint arXiv:2409.12454}, 2024.

\bibitem[Smith et~al.(2014)Smith, Weaver, Grabowski, Rao, and Darvas]{smith2014non}
Smith, M.~M., Weaver, K.~E., Grabowski, T.~J., Rao, R.~P., and Darvas, F.
\newblock Non-invasive detection of high gamma band activity during motor imagery.
\newblock \emph{Frontiers in human neuroscience}, 8:\penalty0 817, 2014.

\bibitem[Song et~al.(2022)Song, Zheng, Liu, and Gao]{song2022eegconformer}
Song, Y., Zheng, Q., Liu, B., and Gao, X.
\newblock Eeg conformer: Convolutional transformer for eeg decoding and visualization.
\newblock \emph{IEEE Transactions on Neural Systems and Rehabilitation Engineering}, 31:\penalty0 710--719, 2022.

\bibitem[Stam et~al.(2007)Stam, Nolte, and Daffertshofer]{stam2007phase}
Stam, C.~J., Nolte, G., and Daffertshofer, A.
\newblock Phase lag index: assessment of functional connectivity from multi channel eeg and meg with diminished bias from common sources.
\newblock \emph{Human brain mapping}, 28\penalty0 (11):\penalty0 1178--1193, 2007.

\bibitem[Su et~al.(2024)Su, Ahmed, Lu, Pan, Bo, and Liu]{su2024roformer}
Su, J., Ahmed, M., Lu, Y., Pan, S., Bo, W., and Liu, Y.
\newblock Roformer: Enhanced transformer with rotary position embedding.
\newblock \emph{Neurocomputing}, 568:\penalty0 127063, 2024.

\bibitem[Valderrama \& Sheoran(2025)Valderrama and Sheoran]{valderrama2025identifying}
Valderrama, C.~E. and Sheoran, A.
\newblock Identifying relevant eeg channels for subject-independent emotion recognition using attention network layers.
\newblock \emph{Frontiers in Psychiatry}, 16:\penalty0 1494369, 2025.

\bibitem[Varela et~al.(2001)Varela, Lachaux, Rodriguez, and Martinerie]{varela2001brainweb}
Varela, F., Lachaux, J.-P., Rodriguez, E., and Martinerie, J.
\newblock The brainweb: phase synchronization and large-scale integration.
\newblock \emph{Nature reviews neuroscience}, 2\penalty0 (4):\penalty0 229--239, 2001.

\bibitem[Wang et~al.(2024{\natexlab{a}})Wang, Liu, He, Xu, Ma, and Li]{wang2024eegpt}
Wang, G., Liu, W., He, Y., Xu, C., Ma, L., and Li, H.
\newblock Eegpt: Pretrained transformer for universal and reliable representation of eeg signals.
\newblock \emph{Advances in Neural Information Processing Systems}, 37:\penalty0 39249--39280, 2024{\natexlab{a}}.

\bibitem[Wang et~al.(2024{\natexlab{b}})Wang, Zhao, Luo, Zhou, Jiang, Li, Li, and Pan]{wang2024cbramod}
Wang, J., Zhao, S., Luo, Z., Zhou, Y., Jiang, H., Li, S., Li, T., and Pan, G.
\newblock Cbramod: A criss-cross brain foundation model for eeg decoding.
\newblock \emph{arXiv preprint arXiv:2412.07236}, 2024{\natexlab{b}}.

\bibitem[Woo et~al.(2024)Woo, Liu, Kumar, Xiong, Savarese, and Sahoo]{woo2024moirai}
Woo, G., Liu, C., Kumar, A., Xiong, C., Savarese, S., and Sahoo, D.
\newblock Unified training of universal time series forecasting transformers.
\newblock \emph{International Conference on Machine Learning}, 2024.

\bibitem[Xu et~al.(2020)Xu, Xu, Ke, An, Liu, and Ming]{xu2020cross}
Xu, L., Xu, M., Ke, Y., An, X., Liu, S., and Ming, D.
\newblock Cross-dataset variability problem in eeg decoding with deep learning.
\newblock \emph{Frontiers in human neuroscience}, 14:\penalty0 103, 2020.

\bibitem[Yi et~al.(2023)Yi, Wang, Ren, and Li]{yi2023learningMMM}
Yi, K., Wang, Y., Ren, K., and Li, D.
\newblock Learning topology-agnostic eeg representations with geometry-aware modeling.
\newblock \emph{Advances in Neural Information Processing Systems}, 36:\penalty0 53875--53891, 2023.

\bibitem[Zhang et~al.(2023)Zhang, Yuan, Yang, Chen, Wang, and Li]{zhang2023brant}
Zhang, D., Yuan, Z., Yang, Y., Chen, J., Wang, J., and Li, Y.
\newblock Brant: Foundation model for intracranial neural signal.
\newblock \emph{Advances in Neural Information Processing Systems}, 36:\penalty0 26304--26321, 2023.

\bibitem[Zhang et~al.(2019)Zhang, Yuan, Huang, Zheng, Wu, Xu, and Pan]{zhang2019cyborg}
Zhang, S., Yuan, S., Huang, L., Zheng, X., Wu, Z., Xu, K., and Pan, G.
\newblock Human mind control of rat cyborg’s continuous locomotion with wireless brain-to-brain interface.
\newblock \emph{Scientific reports}, 9\penalty0 (1):\penalty0 1321, 2019.

\end{thebibliography}

\newpage
\appendix
\onecolumn
\section{Implementation Details}
\subsection{finetuning details}
\label{append_implementation_detail}
\begin{table}[H]
\centering
\caption{Hyperparameters for DIVER fine-tuning.}
\label{hyperparams_CBraMod}
\begin{tabular}{ll}
\toprule
\textbf{Hyperparameters} & \textbf{Settings} \\
\midrule
Epochs & 50 \\
Batch size & 64 \\
Dropout & 0.1 \\
Optimizer & AdamW \\
Learning rate & 1e-4 \\
Adam $\beta$ & (0.9, 0.999) \\
Adam $\epsilon$ & 1e-8 \\
Weight decay & 5e-2 \\
Scheduler & CosineAnnealingLR \\
Cosine cycle epochs & 50 \\
Minimal learning rate & 1e-6 \\
Clipping gradient norm & 1 \\
Label smoothing (multi-class classification) & 0.1 \\
\bottomrule
\end{tabular}
\end{table}

By default, we used the same hyperparameters(\cref{hyperparams_CBraMod}) as CBraMod \cite{wang2024cbramod} in component  ablation analysis (\cref{table2}) and channel permutation(\cref{table3}) studies. 
Some of the cases in component analysis and channel permutation analysis did not show convergence, due to high learning rate.
Trials that didn't converge are listed below.
\begin{itemize}
    \item w/o patch-wise CNN encoding with random seed 42
    \item w/o sepctral embedding with random seed 44
    \item w/o binary attention bias with random seed 44 in pretrain permute and finetuning naive condition 
\end{itemize}
Learning rate was set as 5e-5 only for these trials to keep the consistency of using same random seeds of 41, 42, 43, 44, 45 for all other studies while following reported finetuning settings of CBraMod\cite{wang2024cbramod}.

\section{Reproducibility Problems in \citep{wang2024cbramod}}\label{cbramod reproduction problem}

\begin{table*}[h]
\centering
\footnotesize
\setlength{\tabcolsep}{4pt}
\caption{The results of different methods on downstream tasks tested on our setting}
\label{append_repr_ours}
\begin{tabular}{lcccccc}
\toprule
 & \multicolumn{3}{c}{\textbf{FACED, 9-class}} & \multicolumn{3}{c}{\textbf{PhsyioNet-MI, 4-class}} \\
\cmidrule(lr){2-4} \cmidrule(lr){5-7}
\textbf{Methods} & \textbf{Bal. Acc.(\%)} & \textbf{Kappa(\%)} & \textbf{F1(\%)} & \textbf{Bal. Acc.(\%)} & \textbf{Kappa(\%)} & \textbf{F1(\%)} \\
\midrule
EEGNet              & 16.3 $\pm$ 2.9 & 6.0 $\pm$ 3.4 & 10.6 $\pm$ 3.0 & 57.4 $\pm$ 0.8 & 43.2 $\pm$ 1.1 & 57.3 $\pm$ 0.8 \\
SPaRCNet            & 15.3 $\pm$ 0.6 & 4.9 $\pm$ 0.7 & 9.6 $\pm$ 1.2 & 56.8 $\pm$ 0.7 & 42.4 $\pm$ 1.0 & 56.7 $\pm$ 0.9 \\
ST-Transformer      & 21.9 $\pm$ 0.6 & 12.1 $\pm$ 0.7 & 21.0 $\pm$ 0.7 & 59.5 $\pm$ 0.8 & 45.9 $\pm$ 1.1 & 59.2 $\pm$ 0.9 \\
EEGConformer        & 45.1 $\pm$ 1.2 & 37.9 $\pm$ 1.4 & 45.2 $\pm$ 1.3 & 57.4 $\pm$ 0.5 & 43.2 $\pm$ 0.6 & 57.4 $\pm$ 0.4 \\
\midrule
LaBraM-Base         & 46.6 $\pm$ 1.1 & 39.6 $\pm$ 1.2 & 46.4 $\pm$ 1.1 & 64.7 $\pm$ 0.7 & 52.9 $\pm$ 0.9 & 64.8 $\pm$ 0.8 \\
CBraMod 10\%   & 56.5 $\pm$ 0.8 & 51.0 $\pm$ 1.0 & 56.9 $\pm$ 0.8 & 62.4 $\pm$ 0.6 & 49.9 $\pm$ 0.8 & 62.6 $\pm$ 0.7 \\
CBraMod             & 56.6 $\pm$ 1.2 & 50.9 $\pm$ 1.3 & 56.8 $\pm$ 1.2 & 61.2 $\pm$ 0.8 & 48.4 $\pm$ 1.0 & 61.4 $\pm$ 0.8 \\
\bottomrule
\end{tabular}
\end{table*}

\begin{table*}[h]
\footnotesize
\centering
\setlength{\tabcolsep}{4pt}
\caption{Reported performance of downstream EEG tasks in CBraMod}
\label{append_repr_reported}
\begin{tabular}{lcccccc}
\toprule
 & \multicolumn{3}{c}{\textbf{FACED, 9-class}} & \multicolumn{3}{c}{\textbf{PhsyioNet-MI, 4-class}} \\
\cmidrule(lr){2-4} \cmidrule(lr){5-7}
\textbf{Methods} & \textbf{Bal. Acc.(\%)} & \textbf{Kappa(\%)} & \textbf{F1(\%)} & \textbf{Bal. Acc.(\%)} & \textbf{Kappa(\%)} & \textbf{F1(\%)} \\
\midrule
EEGNet              
& 40.9 $\pm$ 1.2 & 33.4 $\pm$ 2.5 & 41.2 $\pm$ 1.4 
& 58.1 $\pm$ 1.3 & 44.7 $\pm$ 2.0 & 58.0 $\pm$ 1.2 \\
SPaRCNet            
& 46.7 $\pm$ 1.6 & 39.8 $\pm$ 2.9 & 47.3 $\pm$ 1.3 
& 59.3 $\pm$ 1.5 & 45.6 $\pm$ 2.3 & 59.4 $\pm$ 1.5 \\
ST-Transformer      
& 48.1 $\pm$ 0.8 & 41.4 $\pm$ 1.3 & 48.0 $\pm$ 1.0 
& 60.4 $\pm$ 0.8 & 47.1 $\pm$ 2.0 & 60.5 $\pm$ 0.8 \\
EEGConformer        
& 45.6 $\pm$ 1.3 & 38.6 $\pm$ 1.9 & 45.1 $\pm$ 1.1
& 60.5 $\pm$ 1.0 & 47.4 $\pm$ 1.7 & 60.6 $\pm$ 1.0 \\
\midrule
LaBraM-Base         
& 52.7 $\pm$ 1.1 & 47.0 $\pm$ 1.9 & 52.9 $\pm$ 1.0 
& 61.7 $\pm$ 1.2 & 49.1 $\pm$ 1.9 & 61.8 $\pm$ 1.4 \\
CBraMod             
& 55.1 $\pm$ 0.9 & 50.4 $\pm$ 1.2 & 56.2 $\pm$ 0.9 
& 64.2 $\pm$ 0.9 & 52.2 $\pm$ 1.7 & 64.3 $\pm$ 1.0 \\
\bottomrule
\end{tabular}
\end{table*}

In order to compare and evaluate our results with previous studies, we tried to reproduce the performance of prior models. \cref{append_repr_reported} shows performance of downstream tasks reported in CBraMod and \cref{append_repr_ours} shows results of testing on our setting. Though we used the codes and model weights of the previous work, we were unable to replicate the reported results. Learning rate and weight decay used for the downstream task in previous baseline test are not disclosed. Due to these limitations, we cannot exactly yield the reported results. This might arise due to the selection of random seeds, which were also not disclosed by authors. 
We utilized the identical code provided in github, ranging from preprocessing, pretraining and finetuning. 
Even when using opened pretrained weight from CBraMod produced results outside standard deviation for PhysioNet-MI while FACED resulted in a slightly better performance.

\end{document}